\documentclass{article}

\usepackage[english]{babel}

\usepackage[letterpaper,top=2cm,bottom=2cm,left=3cm,right=3cm,marginparwidth=1.75cm]{geometry}

\usepackage{amsmath}
\usepackage{graphicx}
\usepackage{xcolor}
\usepackage{multirow}
\usepackage{array}    
\usepackage[colorlinks=true, allcolors=blue]{hyperref}
\usepackage[backend=bibtex,style=ieee,natbib=true,sorting=none,url=false,doi=false,isbn=false]{biblatex} 
\AtBeginBibliography{\small}
\title{Dynamic neuronal networks efficiently achieve classification in robotic interactions with real-world objects}

\author{Pakorn Uttayopas, Xiaoxiao Cheng, Udaya Bhaskar Rongala,\\ Henrik J\"{o}rntell, Etienne Burdet}
\date{}
\begin{document}
\maketitle

\begin{abstract}
Biological cortical networks are potentially fully recurrent networks without any distinct output layer, where recognition may instead rely on the distribution of activity across its neurons. Because such biological networks can have rich dynamics, they are well-designed to cope with dynamical interactions of the types that occur in nature, while traditional machine learning networks may struggle to make sense of such data. Here we connected a simple model neuronal network (based on the 'linear summation neuron model' featuring biologically realistic dynamics (LSM), consisting of 10 of excitatory and 10 inhibitory neurons, randomly connected) to a robot finger with multiple types of force sensors when interacting with materials of different levels of compliance. Scope: to explore the performance of the network on classification accuracy. Therefore, we compared the performance of the network output with principal component analysis of statistical features of the sensory data as well as its mechanical properties. Remarkably, even though the LSM was a very small and untrained network, and merely designed to provide rich internal network dynamics while the neuron model itself was highly simplified, we found that the LSM outperformed these other statistical approaches in terms of accuracy.
\end{abstract}






\section{Introduction}
Here we aimed to use biologically relevant neuron models connected in a brain-like network structure to study its potential to achieve input separation in a robotic system interacting with real-world objects. The model network was inspired by local cortical networks in its recursive structure in principle, though with much fewer neurons and without the ambition to precisely mimick any assumed specific network structure. The aim was to explore if the inherent dynamic properties in such networks in themselves were enough to achieve efficient object classification.

Our model system is reminiscent of Reservoir Computing networks (i.e. Gauthier et al 2020 Nature Communications), but our neurons have state memory, i.e. dynamics, which are biologically relevant. Moreover, the population of neurons are split into excitatory and inhibitory neurons. Combined with the neuronal output thresholding, i.e. imparting nonlinearity to the networks when inhibition drives the neurons below their thresholds, and combined with biologically relevant conduction delays, this setting creates extraordinarily rich network dynamics.

Motivation for: what would be required in the robotics design to explore the questions we set out to explore? How well could we live up to those requirements with the robotics system used?

\section{Methods}

\subsection{Neuron model}
The neuron model used in this work was a non-spiking Linear Summation Model (LSM) with an additional dynamic leak component \cite{LSM}. LSM aims to capture the important characteristics of a Hodgkin-Huxley (H-H) conductance-based model \cite{HHmodel}. The LSM describes the activity dynamics $\{a_j(t)\}$ of ``cortical neurons'' arising from the weighted activity of other cortical neurons (with inhibitory $\alpha_i<0$ and excitatory $\alpha_i>0$ projection) as well as sensory neurons $\{b_k\}$:
\begin{equation}
\tau \,\frac{da_j}{dt} = \, - a_j^+(t) + \frac{\sum_{i \neq j} \alpha_i\,a_i(t) + \sum_k \beta_k\,b_k(t)}{\sum_{i \neq j} |\alpha_i|\,a_i(t) + \sum_k |\beta_k|\,b_k(t) + \gamma} \, , \quad 
a_j^+ = \begin{cases} a_j \quad a_j>0 & \\
0  \quad a_j \leq 0 &
 \end{cases}
\label{eq:MI}
\end{equation}
Importantly, the forgetting of each cortical neuron activity $-a_j^{+}(t)$ ensures convergence, at a speed regulated by the time constant $\tau$. The denominator weights the influence of other neurons, with a factor $\gamma>0$ avoiding divergence when they are not active.

\subsection{Neuron models configuration}
\subsubsection{Network Connectivity}
\begin{figure}[h]
\centering
\qquad \qquad
\includegraphics[width=1\columnwidth]{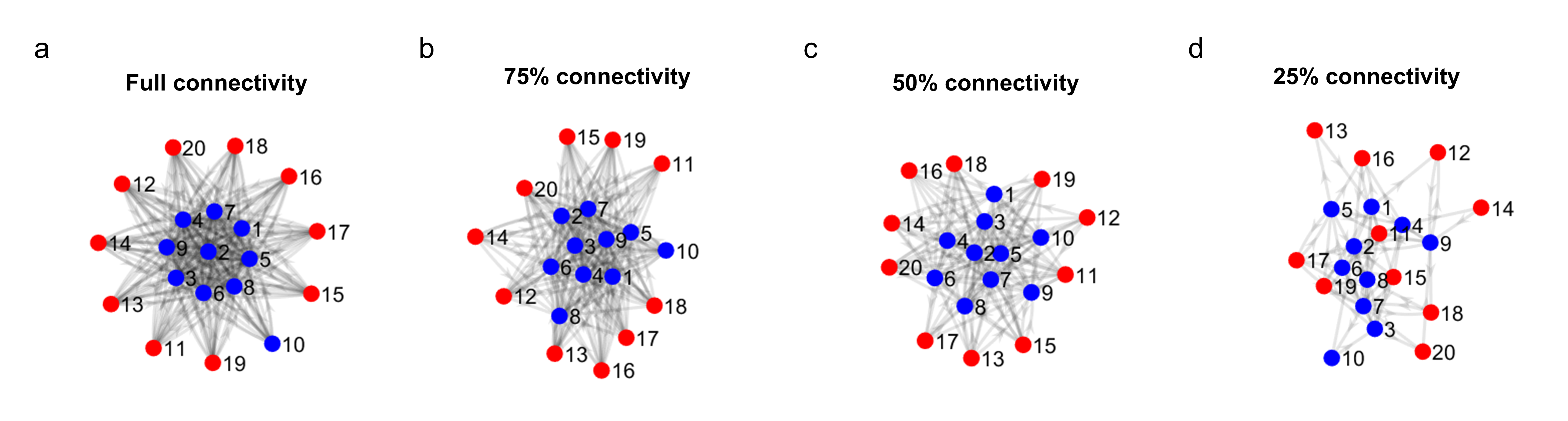}
\caption[]{Visualization of the network connections at the different connectivity levels. In (a)-(d), red nodes are inhibitory neurons and blue nodes are excitatory neurons. Full connectivity (a) implied that every single node (neuron) was connected to all other nodes, but the weight of each connection was varied according to random patterns. Reduced connectivity was achieved by removing  connections randomly to (b) 75\% (c) 50\% (c) 25\% connectivity.}
\label{fig:connectivity}
\end{figure}

In this work, the network connectivity is comprised of 10 excitatotry nodes (ENs, blue markers) and 10 inhibitory nodes (INs, red markers), totalling 20 neurons. All nodes randomly received the external inputs which are force, torque and vibration from object-robot interaction. ENs are bi-directionally connected with all other nodes, but the central one. However, INs are connected only with ENs, but not within INs (see Fig \ref{fig:connectivity}.).

4 connectivity configurations \{full connectivity, 75\% connectivity, 50\% connectivity and 25\% connectivity\} were considered. In full connectivity, all nodes within the network are fully connected as mentioned previously (see Fig \ref{fig:connectivity}.a). In other configurations, we randomly removed 25\%, 50\% and 75\% connections from the full connectivity (see Fig \ref{fig:connectivity}.b-d respectively). Note that full connectivity was used for the rest of network simulations, otherwise specified.

\subsubsection{Static and Dynamic leak}
The LSM neurons have both a static leak and a dynamic leak constant \cite{LSM}. The static leak constant, $k$ was fixed at $2$ as it only acts as a normalization factor of the neuron activity, without affecting the network dynamics. On the other hand, we consider the dynamic leak time constant at $\tau_{Dyn}=1/100$ as this constant acts as a low-pass filtering factor for neuron activity and thereby impact the network dynamics. 

\subsubsection{Conduction Delays}
 An axonal delays between two neurons were also applied in the network. The conduction delays  in communication among neurons were uniformly distributed randomised in range of [0,1] ms. These values of conduction delays in the simulations denotes as timesteps of signal lagged between two neurons in the network. 

\subsubsection{Input Weights}
All external inputs from ENs are weighted. The weights were generated as normal distributions with a distribution mean ($\mu$) of 0.3 and standard deviation ($\sigma$) of 0.5. 

\subsubsection{Synaptic Weights}
All connections in the network were weighted. The weights were generated as normal distribution with a distribution mean ($\mu$) of 0.1 and standard deviation ($\sigma$) of 0.05. The weights of INs outgoing connections were set negative and the ENs were made positive.

\subsection{Metrics}
\subsubsection{Information gain}
The information gain (IG) can be used to measure the uncertainty in prediction of objects when observation or sets of feature is introduced \cite{appliedInformationgain}. This information gain is defined as 

\begin{equation}
I(C;X) = \, H(C)-H(C|X) 
\label{eq:MI}
\end{equation}
where $H(C)$ is an entropy of random variable representing the object classes, $C$ and $H(C|X)$ is an entropy of the object classes, $C$ given that the knowledge of a random variable representing an acquired observation, $X$. 

In this work, object classes are discrete variables and the acquiredfeatures are continuous variables. The information gain can be rewritten as 

\begin{equation}
I(C;X) = \sum_{c}^{C}\! P(c)\int_{x}\,P(x|c)\,\ln \frac{P(x|c)}{P(x)}\,dx \
\end{equation}
where $P(c)$ represents the probability distribution of an object class $C$, $P(x|c)$ is the conditional probability of an observed feature $x$ given object class $c$ and $p(x)$
is the probability density of  an observed feature $x$. This information gain was computed for each set of feature, a higher value indicate more uncertainty is removed from the prediction when that set of feature is used as the features in a classifier.

where $H(C)$ is an entropy of random variable representing the object classes, $C$ calculated by $H(C) = \, -\sum_{\substack{c\in C}}\ \!\! P(C) \, \log P(C) \,$, and $H(C|X)$ is an entropy of the object classes, $C$ given that the knowledge of a random variable representing an acquired observation, $X$ calculated by $H(C|X)= \, -\sum_{\substack{c\in C}}\sum_{\substack{x\in X}}\ \!\! P(c,x)\, \log P(c|x) \,.$




\subsection{Experimental set up}

\begin{figure}[tb]
\centering
\qquad \qquad
\includegraphics[width=0.75\columnwidth]{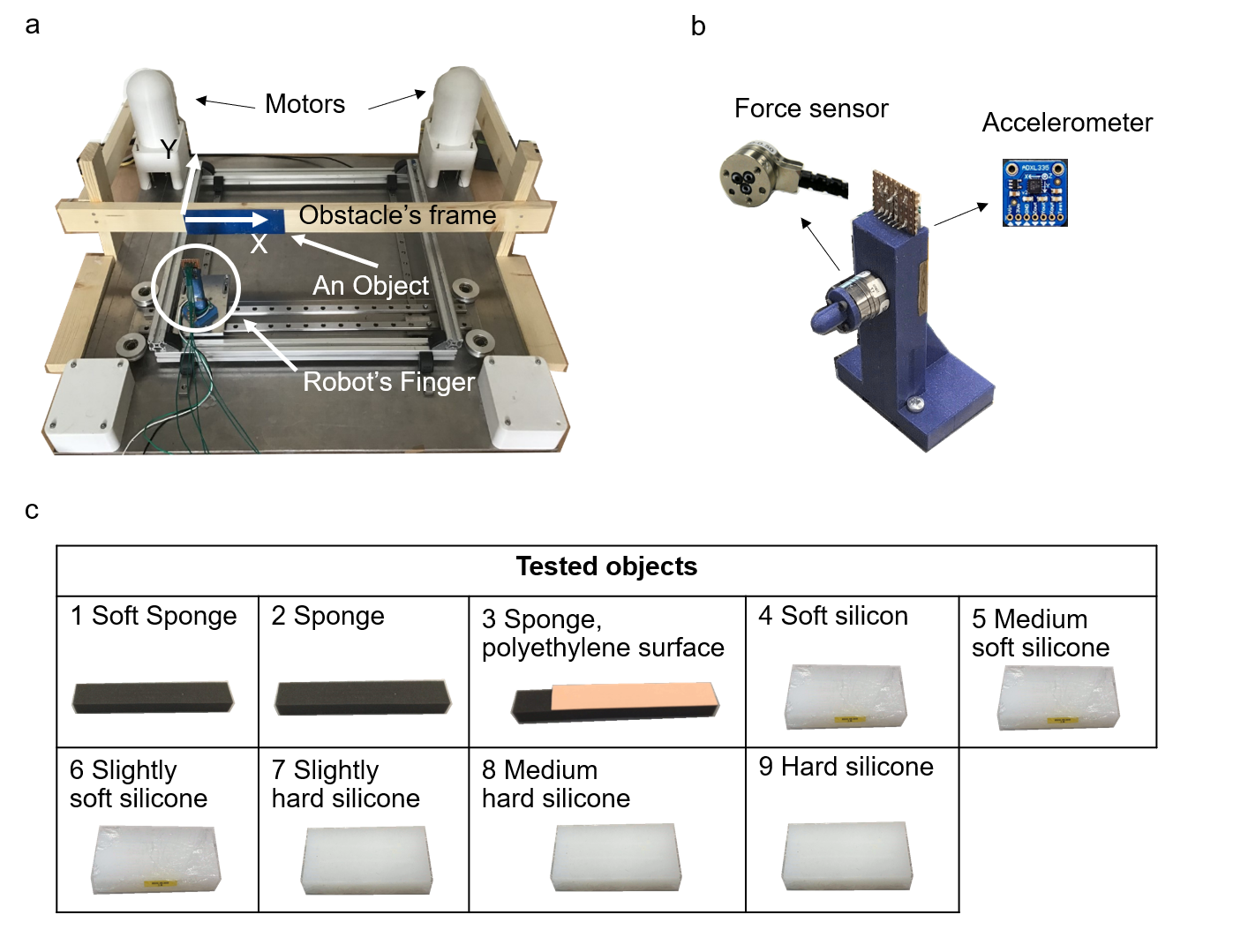}

\caption[]{Experiment. a. HMan with a finger and an object. A wooden frame is used to attach the object (blue area) for the robot to explore. The finger is driven by two motors in 2D planar where the force sensor is always faced toward the object’s surface. b. A 3D printed robot’s finger with force/torque sensor attached on its front and accelerometer on its back. c. Objects used in the experiment}
\label{fig:H-man}
\end{figure}

The HMan robot \cite{H-manrobot} shown in Fig.\,\ref{fig:H-man}.a was used to haptically interact with objects to gather tactile information. This planar robotic interface with two degrees of freedom is equipped with a six-axis force/torque sensor (model SI-25-0.25; ATI Industrial Automation) between the tip and base of a robot's finger to measure interaction force along with interaction torque. Besides, a three-axis accelerometer (model ADXL335; Analog Devices) is mounted on the back of robot's finger to measure its vibration. The robot's finger can be seen in Fig.\,\ref{fig:H-man}.b .

The robot interacted with 9 different \textbf{soft} objects shown in Fig.\,\ref{fig:H-man}.c: \{soft sponge, sponge,sponge with polyethylene foam surface, soft silicon, medium soft silicone, slightly soft silicone, slightly hard silicone, medium hard silicone and hard silicone\}. It uses one main actions to interact which is \textit{indentation}. The robot moves its finger on the object's surface toward the normal direction with a desired trajectory $D_{r}=d\sin(\pi\tau)+d$ $m$, where $d$ is a desired depth and $\tau$ is a time interval as listed in Table \ref{table:action}. Tactile information in the form of interaction force, torque and vibration was collected for 25 times for each pair of sub-action and object.

\begin{table}[h]
\caption{Actions used in the experiment} 
\label{table:action}
\begin{center}
\begin{tabular}{ |c|c|c|c| } 
\hline
Action & Sub action & Depth $(m)$ & Time $(s)$\\
\hline
Indentation & \begin{tabular}[c]{@{}l@{}} Light\\ Medium \\ Deep\end{tabular} & \begin{tabular}[c]{@{}l@{}} $d=0.0075$\\$d=0.01$  \\ $d=0.015$ \end{tabular} & $\tau =(0,20]$ \\ 

\hline
\end{tabular}
\end{center}
\end{table}

\subsection{Data processing}

As the robot pressed its finger against material's surface in the normal direction, only normal force, normal vibration and 3 directions torque showed responses from this interaction. Thus, we have considered using them in the LSM network to avoid bias from non-activated sensory inputs. This information will be referred as ``\textit{raw data}".

The raw data were processed to extract features. Basic temporal statistical features including  maximum values, mean values, and standard deviation values from 8 sensors were extracted. In total, 26 ``statistical features" were extracted from the tactile information. 

This raw data is also used to extract mechanical properties of the objects such as stiffness and viscosity. These stiffness and viscosity values were obtained by estimating parameters in the linear spring-damper model with interaction force and robot positions using the least squares method.  

Additionally, the raw data was fed to the LSM network to obtain its outputs. The same basic temporal statistical features; mean values, and standard deviation values were also extracted from all 20 LSM outputs. In total, 60 ``LSM features" are obtained.

All features are then used to form sets of feature. The statistical features and mechanical properties were combined to form sets of feature used as baselines to compare with the results obtained from the LSM features. In addition, to avoid overfitting when the classification is performed, a common dimensionality reduction technique, principle component analysis (PCA), was applied to those sets of features before feeding them into a classifier. Table \ref{table:sum_set_features} show a summary of sets of feature.

\begin{table}[h]
\caption{4 sets of feature used for comparing object recognition results.}
\label{table:sum_set_features}
\begin{center}
\begin{tabular}{|l|l|}
\cline{1-2}
\multicolumn{1}{|c|}{\textbf{Denomination}} & \multicolumn{1}{c|}{\textbf{Features}} \\ \cline{1-2}

FTV-pca &  \begin{tabular}[c]{@{}l@{}} 10 principle components \\ from statistical features \end{tabular}     \\\cline{1-2}

FTVSV-pca &  \begin{tabular}[c]{@{}l@{}} 10 principle components \\ from statistical features, \\ stiffness and viscosity \end{tabular}     \\\cline{1-2}

LSM-pca &  \begin{tabular}[c]{@{}l@{}} 10 principle components\\ from LSM features \end{tabular}     \\\cline{1-2}

LSMex-pca &  \begin{tabular}[c]{@{}l@{}} 10 principle components \\ from LSM features, \\ only ENs are sensory input node \end{tabular}     \\\cline{1-2}

\end{tabular}
\end{center}
\end{table}

\subsection{Classifier}
In view of the uncertain nature of the measurements we propose using a Bayesian classifier to identify objects. In particular, the Naive Bayes classifier calculates the posterior probability of an object class $c$ given features $X$ in a feature set using Bayes's Theorem:
\begin{equation}
P(c|X) = \frac{P(X|c)\,P(c)}{P(X)}
\label{eq:baye}
\end{equation}
where $P(X|c)$ is a likelihood of features $X$ given object class $c$, $P(c)$ is a prior belief of class $c$ and $P(X)$ is a probability of features $X$. The predicted object class \(c^*\) from the acquired features $X$ is the class $c$ that maximises the posterior probability $P(c|X)$:
\begin{equation}
c^*= \arg\max_{c \epsilon C} P(c|X) \, 
\label{eq:class_baye}
\end{equation}
The Bayesian classifier were evaluated with 50:50 ratio of training and testing set of feature for 100 repetitions to obtain results for each set of feature.

\section{Results}
\subsection{LSM responses}
\label{section:LSM_results}

\begin{figure}[h]
\centering
\includegraphics[width=0.75\columnwidth]{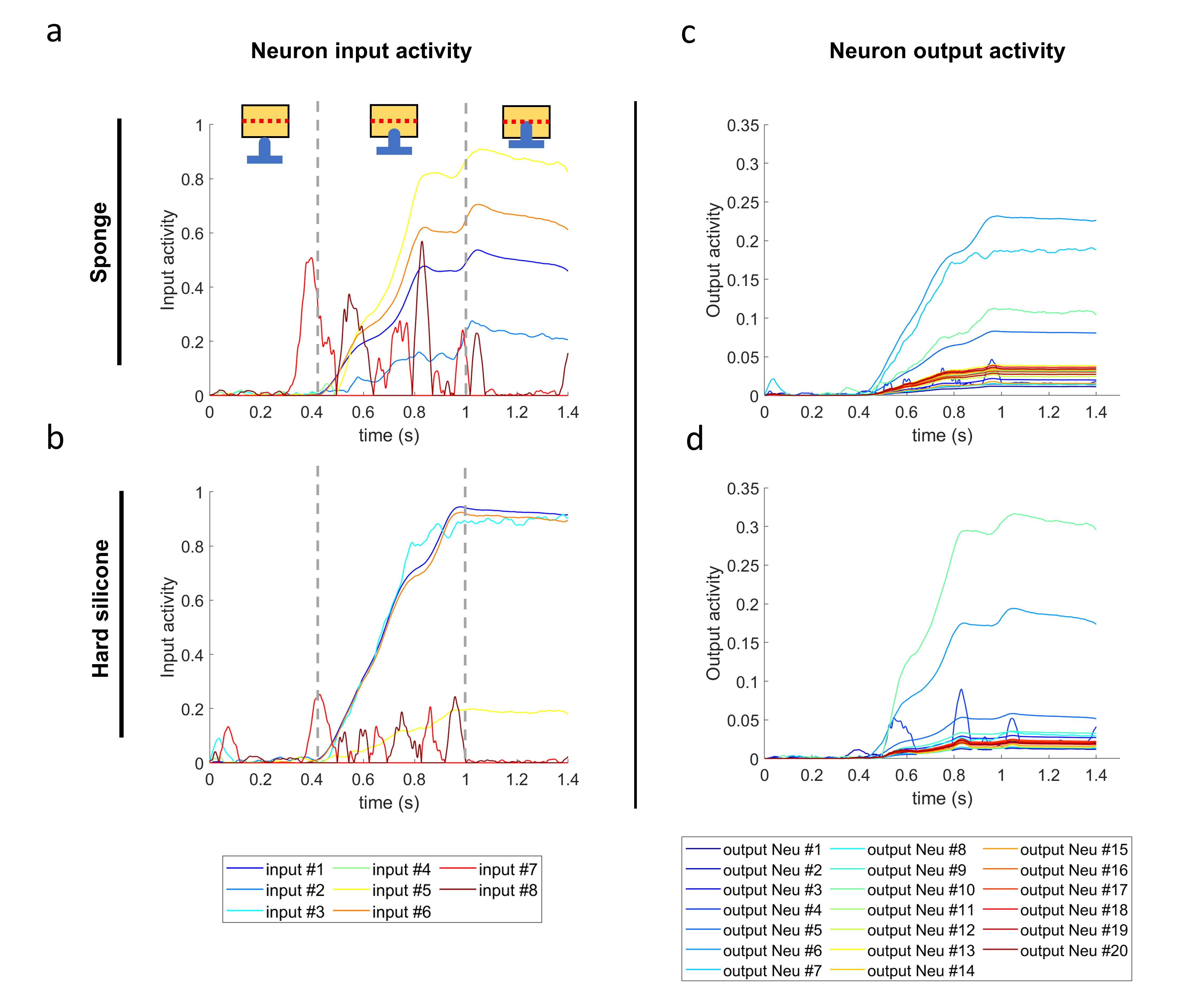}
\caption[]{LSM network responses to 2 given inputs from sponge and hard silicone. (a)-(b) The neuron input activity given by a sponge and hard silicone with deep indentation where indicators of the robot finger position relative to an object are shown. (c)-(d) The neuron output activity from the neuron network, with the given inputs from (a) and (b) respectively.}
\label{fig:example_LSM_response}
\end{figure}

As the robot was commanded to move its finger against different materials, the information from the robotic sensors were fed to our LSM network to drive activity within that network. Figure \ref{fig:example_LSM_response} shows the sensor inputs and the resulting activity across the neuronal population for two sample materials using in full network connectivity. Whereas the sensors had a baseline activity without bias, all neurons started out with 0 activity (Fig. \ref{fig:example_LSM_response}). When the robot was commanded to move against the materials at time 0.4 s, the output activities of the neurons start to built up due to the sensor input activity. The neuron activities were then allowed to reach a steady state at 1 s as the robot have reached the maximum deformation point of the materials. The sensor input in turn drove an evolving activity of the neurons, which depended both on the sensor input and on the internal connectivity and the internal dynamics of the neuronal network.

\subsection{Classification Comparison}
\label{section:classification_results}

\begin{figure*}[!tb]
\centering
\qquad \qquad
\includegraphics[width=1\columnwidth]{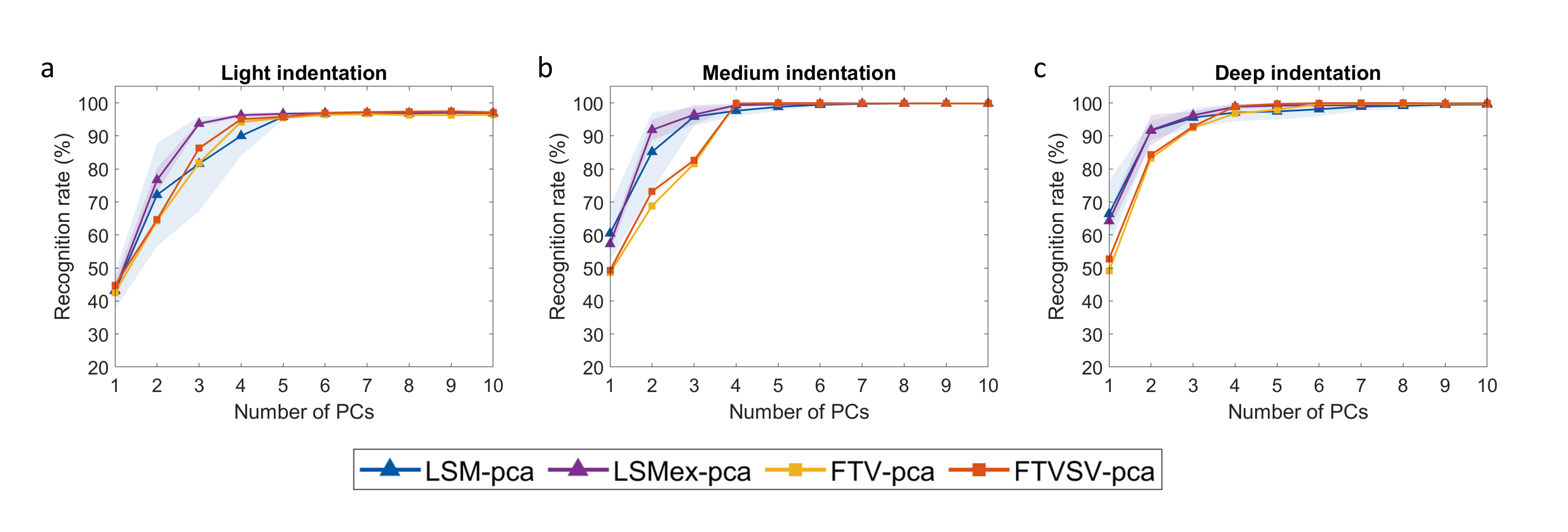}
\caption[]{Comparison of PC-based classification performance for sensory outputs versus the LSM output. (a)-(c) Recognition rate and variance for three different indentation levels as a function of the number of PCs. (d)-(f) Information gain for the same conditions.}
\label{fig:acc_results}
\end{figure*}

 Fig \ref{fig:acc_results}. shows the recognition rate for each set of features as a function of number of PCs as well as their standard deviation. LSM-pca provided the same recognition rate as others for light indentation levels. However, for medium and deep indentation levels, LSM-pca  provided higher rate than the other sets of PCs feature provide especially in range of 1 to 4 PCs size. Unsurprisingly, when using only ENs as sensory inputs nodes, the LSM network provided higher recognition rate than other sets of feature in 1-4 PCs size even in the light indentation levels.


\subsection{Effects of sensory inputs types}

\begin{figure}[!tb]
\centering
\qquad \qquad
\includegraphics[width=1\columnwidth]{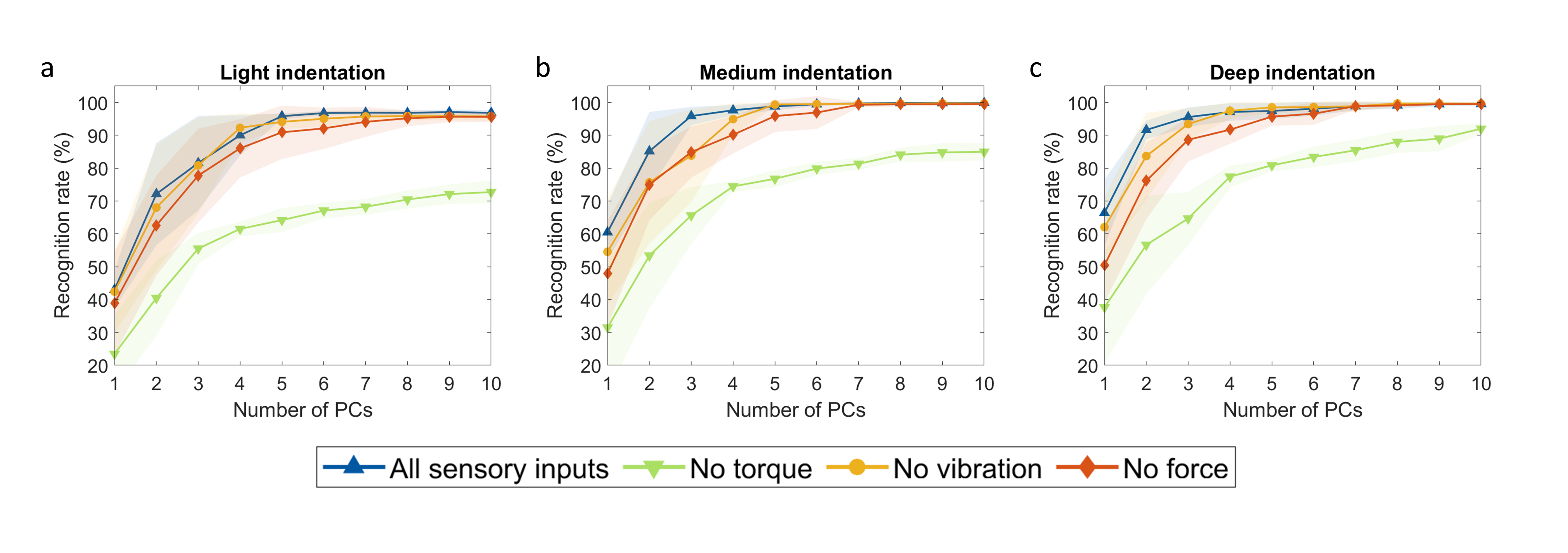}
\caption[]{Performance of the LSM network a type of sensory inputs is ignored at the time. (a) Percentage of recognized materials as a function of the number of PCs from the LSM network for light indentation. (b) Similar display as in (a) but for medium indentation. (c) As in (a) and (b), but for deep indentation.}
\label{fig:acc_sensory}
\end{figure}

Fig. \ref{fig:acc_sensory}. shows recognition rates obtained by using neuron responses with different combinations of sensory input types. Unsurprisingly, providing all types of sensor: \{force, torque, vibration\} into the LSM network yielded the highest results in all indention levels. By dropping vibration information, recognition rated decreased. However, their values were higher than results from dropping fore information from the sensory inputs especially in 1-7 PCs size. Lastly, by ignoring torque information in the system, recognition rates become the lowest results comparing with others cases within the same indention levels.

\subsection{Effects of Connectivity}
\begin{figure}[!tb]
\centering
\qquad \qquad
\includegraphics[width=1\columnwidth]{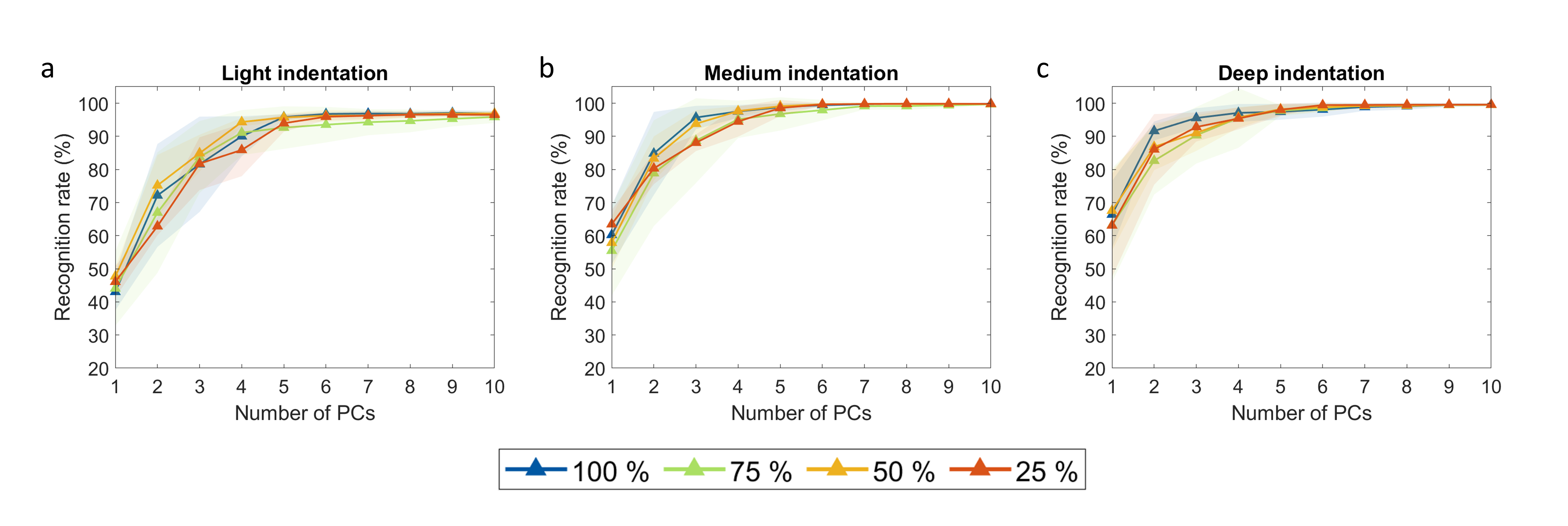}
\caption[]{Performance of the LSM network at different levels of connectivity. (a) Percentage of recognised materials as a function of the number of PCs from the LSM network for light indentation. (b) Similar display as in (a) but for medium indentation. (c) As in (a) and (b), but for deep indentation.}
\label{fig:acc_connection}
\end{figure}

Fig. \ref{fig:acc_connection}. shows recognition rates from different connectivity configurations of the LSM network as a function of number of PCs. The recognition rates for all connectivity configurations are roughly the same in all indention levels. Slight effects of the network connectivity can be found especially in 1-4 PCs size from medium and deep indention levels. The recognition rate decreased when the connectivity are reduced from full to 75\%. As network connectivity reduce to 50\%, its recognition rate turn out to be roughly the same as in full connectivity for medium indentation levels. At 25 \% connectivity, the LSM provide lowest recognition rate in light and medium indentation levels.

\section{Discussion}
\subsection{Effects of indentation}
\begin{figure}[h]
\centering
\qquad \qquad
\includegraphics[width=1\columnwidth]{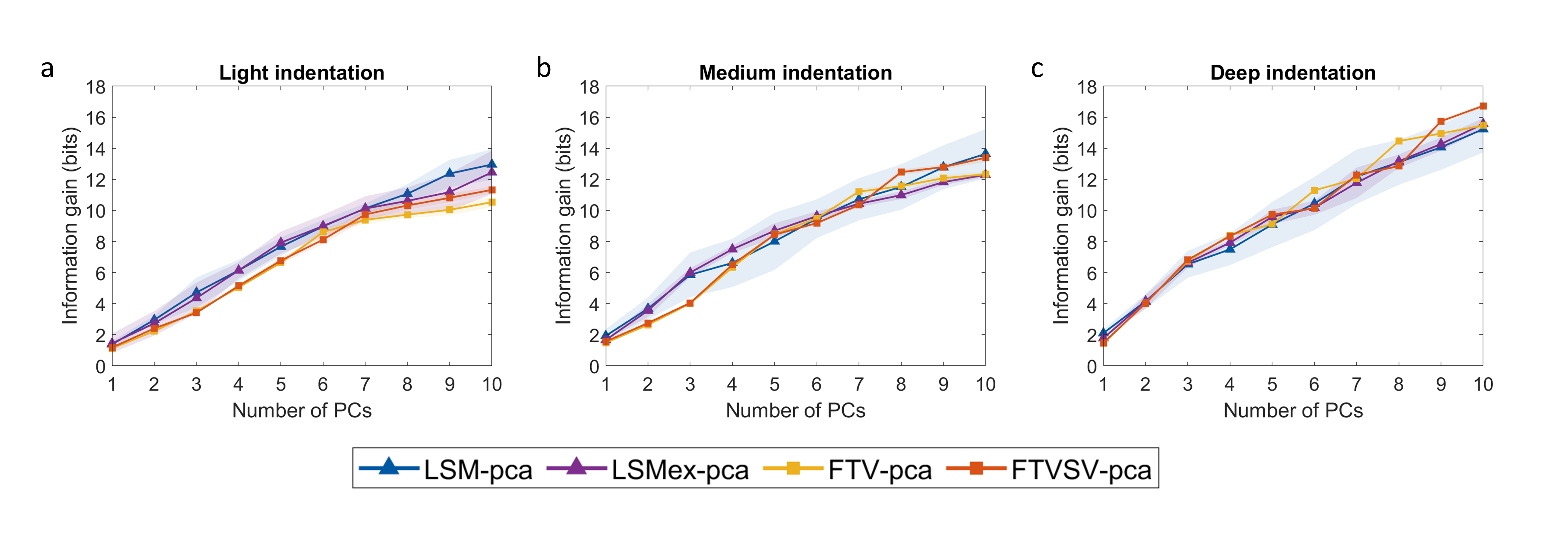}
\caption[]{Comparison of information gain from PC-based sensory outputs versus the PC-based LSM output (a)-(c) Information gain for three different indentation levels as a function of the number of PCs.}
\label{fig:acc_Informationgain}
\end{figure}

According to the classification results, the recognition rate increased as the level of indention increased. Beside, the deeper indentation, the fewer number of PCs used to reach their saturate points. This might be the fact that deeper indentation provides higher amplitude values of interaction force, torque and vibration. Thus, more informative data used to distinguish different objects can be achieved (see Fig. \ref{fig:acc_Informationgain}.a-c for light, medium and deep indentation levels respectively). Even though the deepest level of indentation provided better results, it could damage the objects. Optimal strategies to choose indentation depth to maximise information as well as minimise object damage might be needed to provide the best results in real world applications. However, this is out of the scope for this work. 


\subsection{LSM vs statically tools and mechanical properties}

Using PC-based neuron output activity as features yield the best classification results in term of accuracy especially when only the ENs received the sensory inputs. To find out the reason why, the information gain from sets of PC-based feature in all depth conditions is investigated. 

Fig \ref{fig:acc_Informationgain} shows the calculated information gain as a function of number of PCs for each set of PC-based features. Information gain of PC-based from both LSM and LSMex were slightly higher than results from PC-based statistical features and mechanical properties in all PCs size for light indentation, but in 1-5 PCs size for medium indentation. These results suggests that PC-based LSM outputs could reduced an uncertainly of object prediction better than other sets of feature,by containing additional information. However, it appeared that all sets of feature provided roughly the same information gain in deep indentation. This may be explained by that deep indentation may already provide essential information that LSM can only extract so neglectable information from them.

This additional information might come from the fact that the LSM network acts as the sensory fusion algorithm. The LSM network amplifies information from the sensory inputs by capturing signal dynamics, combining and reflecting them into each neuron activity. These can be clearly observed especially when only the ENs received the sensory inputs. Their classification results become even better than assigning sensory inputs into both INs and ENs. This could be explained by that providing sensory inputs to INs may cause a reduction of activity and thereby information. In constant, providing sensory inputs to ENs amplifies neuron activity, thus more differentiable information can be achieved. However, noises caused by robotics systems can also be amplified in the LSM network causing bias in the system. To avoid such bias, sensory inputs need to be well prepared by filtering unwanted signal.

\subsection{LSM parameters}
\label{discuss:para}

The effects of LSM parameters; \{sensory input types and network connectivity\} on object classification results have been investigated. For sensory input types, the absent of torque largely reduced recognition rates in all indentation levels suggesting that torque information was the most important source of information used to distinguish objects. This could be explained by the fact that 5 out of 8 sensory inputs come from the torque sensor. By dropping out half of sensory inputs, LSM network would lose most of the information resulting in poor classification results. The absences of force and vibration had slightly effects on classification results. The absence of force results in lower recognition rates than the absence of vibration suggesting that force information was more important than vibration information. In fact, at some PCs sizes, the absence of vibration yielded the same recognition rate as all sensory inputs combine.

\begin{figure}[h]
\centering
\qquad \qquad
\includegraphics[width=1\columnwidth]{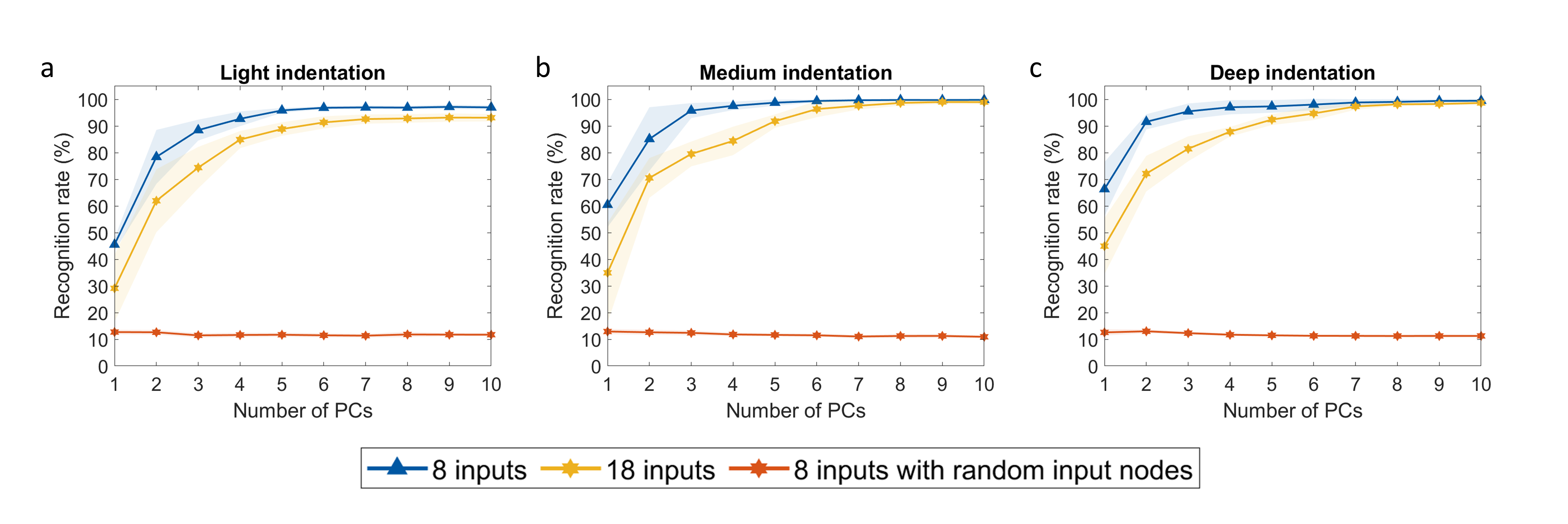}
\caption[]{Performance of the LSM network at different sensory inputs configurations. (a) Percentage of recognised materials as a function of the number of PCs from the LSM network for light indentation. (b) Similar display as in (a) but for medium indentation. (c) As in (a) and (b), but for deep indentation.}
\label{fig:acc_sensory_2}
\end{figure}

2 additional cases of sensory inputs were also investigated which are: (1) using both activated and non-activated sensor responses as sensory inputs (18 sensors) and (2) assigning sensory inputs randomly every repetitions in simulation of the LSM network (8 inputs with random input nodes). Both cases appeared to have lower recognition rates comparing to the results obtained from the defaults setting of the LSM network (see Fig. \ref{fig:acc_sensory_2}). For the first case, this could be explained by that including non-activated sensory inputs may cause bias in the neuron outputs leading to poor classification results. For the second case, the neuron outputs activity may response differently as sensory inputs nodes were randomly assigned. As a results, the classifier could not capture the patterns for each neuron output node leading to the worst recognition rates in this work.

For network connectivity configurations, full connectivity yielded the highest recognition rate comparing with other connectivity configurations. This could be explained by that when the the LSM network was fully connected, providing sensory inputs to the receiving nodes may cause the neuron activity to be fully amplified. In contrast, providing sensory inputs in partial connected nodes of the LSM network may cause less neuron activity, thus less informative data is transferred or amplified to the outputs.

\section{Conclusions}
In this paper, the neuron output activity form the LSM network have been used to recognise 9 soft objects using three different 3 levels of indentation. The classification results showed that using PC-based LSM outputs as features can outperform other sets of features counting raw tactile data and mechanical properties in medium and deep indentation levels. We also found that providing activated sensory inputs only in ENs would improve recognition rate from the defaults setting. The reason behind these results is that the LSM can amplify necessary information used to recognise objects especially when all nodes are fully connected. This can be observed in the information gain as the LSM network provided higher values than other set of features.  However, this study has been only studied with soft objects with indentation.  More objects as well as interactions such as sliding or tapping on material's surfaces might need to be further studied.




\printbibliography[]


\end{document}